\documentclass[ reprint, onecolumn, superscriptaddress, floatfix, amsmath,amssymb, aps ]{revtex4-2}
\usepackage[utf8]{inputenc}
\usepackage[letterpaper,top=2cm,bottom=2cm,left=3cm,right=3cm,marginparwidth=1.75cm]{geometry}

\usepackage{amsmath, amsfonts,amssymb}
\usepackage{physics}
\usepackage{graphicx}
\usepackage[colorlinks=true, allcolors=blue]{hyperref}
\usepackage{dsfont}
\usepackage{hyperref}
\usepackage{mathtools}
\usepackage{braket}
\usepackage{svg}
\usepackage{siunitx}

\renewcommand{\eqref}[1]{Eq.~(\ref{eq:#1})}

\newcommand{\citeasnoun}[1]{Ref.~\cite{#1}}

\newcommand{\Figref}[1]{Figure~\ref{fig:#1}}
\newcommand{\figref}[1]{Fig.~\ref{fig:#1}}
\renewcommand{\eqref}[1]{Eq.~(\ref{eq:#1})}
\newcommand{\Eqref}[1]{Equation~(\ref{eq:#1})}
\newcommand{\eqreftwo}[2]{Eqs.~(\ref{eq:#1},\ref{eq:#2})}

\newcommand*{\mybox}[1]{\framebox{#1}}

\def\XXint#1#2#3{{\setbox0=\hbox{$#1{#2#3}{\int}$}
\vcenter{\hbox{$#2#3$}}\kern-.5\wd0}}

\begin{document}

\title{Approaching upper bounds to resonant nonlinear optical susceptibilities with inverse-designed quantum wells}

\author{Hao Li}
\author{Theodoros T. Koutserimpas}
\author{Francesco Monticone}
\author{Owen D. Miller}
\date{\today}

\maketitle

\textbf{Abstract} \\
We develop a unified framework for identifying bounds to maximum resonant nonlinear optical susceptibilities, and for ``inverse designing'' quantum-well structures that can approach such bounds. In special cases (e.g. second-harmonic generation) we observe that known bounds, a variety of optimal design techniques, and previous experimental measurements nearly coincide. But for many cases (e.g. second-order sum-frequency generation, third-order processes), there is a sizeable gap between the known bounds and previous optimal designs. We sharpen the bounds and use our inverse-design approach across a variety of cases, showing in each one that the inverse-designed QWs can closely approach the bounds. This framework allows for comprehensive understanding of maximum resonant nonlinearities, offering theoretical guidance for materials discovery as well as targets for computational design.

\section*{Introduction}
In this paper, we show that ``inverse-designed'' quantum wells (QWs) can exhibit resonant nonlinear optical susceptibilities that approach fundamental limits for passive nonlinear response. There has been extensive work identifying specific quantum-well structures that exhibit enormous resonant enhancement \cite{lee2014giant,lee2016ultrathin,yu2019third}, and a separate thread of research that has used dipole-matrix-element sum rules to identify upper bounds \cite{kuzyk2000physical,kuzyk2003erratum,kuzyk2000fundamental,kuzyk2003fundamental,kuzyk2006fundamental}, especially for non-resonant nonlinearities in molecules. Here we present a unified general theory for resonant quantum wells, describing bounds for a large class of second- and third-order nonlinearities and demonstrating that computational design techniques can approach them across a wide range of frequencies. In special cases (e.g. second-harmonic generation) we find that bounds from Kuzyk et al. \cite{kuzyk2006fundamental} and computational designs from the literature \cite{radovanovic2001two} already nearly coincide, showing that both are optimal or nearly optimal. For more general cases we derive new bounds, and develop a simple adjoint-gradient-based computational design technique that approaches the bounds. A key utility for bounds is enabling predictions of parameters that can dramatically enhance the ultimate performance limits of a system. We offer two such observations: (1) a 10X reduction of QW effective masses would lead to 32X and 100X enhancements in $\chi^{(2)}$ and $\chi^{(3)}$ nonlinear susceptibilities, respectively, and (2) utilizing transitions amongst higher-order energy levels can yield similar order-of-magnitude susceptibility increases. More generally, sum-rule-based nonlinear susceptibility bounds can be expected to closely predict the maximum nonlinear response of engineered materials. They predict maximal response over all possible geometric designs, for given material parameters, and illuminate how those parameters should be optimized in future synthesis efforts.

\begin{figure*}[tb]
\centering
\includegraphics[width=1\linewidth]{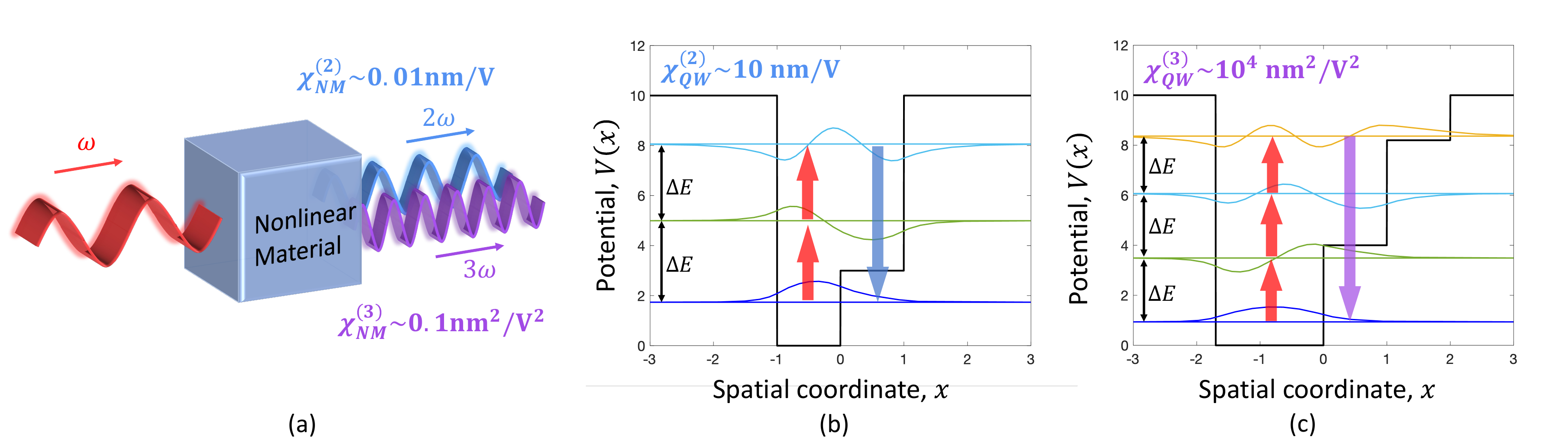}
\caption{\small (a) Conventional bulk nonlinear materials used for second-/third-harmonic generation typically have $\chi^{(2)}\sim\SI{0.01}{nm/V}$, $\chi^{(3)}\sim\SI{0.1}{nm^2/V^2}$, respectively; (b,c) schematic representation of engineered QWs with equally spaced first three/four states, designed for on-resonant SHG and THG processes, yielding high nonlinear susceptibilities: $\chi^{(2)}\sim\SI{10}{nm/V}$, $\chi^{(3)}\sim\SI{e4}{nm^2/V^2}$, respectively.}
\label{fig:setup}
\end{figure*}

Materials with large nonlinear susceptibilities are essential for various fields of science and technology. Typical applications enabled by nonlinear materials include laser technology \cite{boyd2008nonlinear}, frequency combs \cite{ye2005femtosecond,gaeta2019photonic}, 
multi-photon microscopy \cite{diaspro2006multi}, quantum computation and communication \cite{dell2006multiphoton}, etc. Extensive studies have searched for record-high material nonlinearities in organic materials \cite{kaino1993organic}, nonlinear crystals \cite{nikogosyan2006nonlinear}, metal-organic framework \cite{evans2002crystal} and meta-materials \cite{o2015predicting}. However, most conventional materials provide relatively weak bulk nonlinear response, as in \figref{setup}(a). A different approach to achieving large nonlinearities is bandgap engineered semiconductor quantum wells (QWs) \cite{fejer1989observation,rosencher1989second,capasso1994coupled,rosencher1996quantum,gmachl2003optimized}. By tailoring the physical properties of QWs (e.g. well width, barrier height, effective mass, etc.), one can obtain large transition dipole moments ($\sim$ \SI{}{nm}) and controlled energy levels to achieve resonance conditions. This results in a significant enhancement of SHG and THG nonlinear susceptibilities, by $10^3$X and $10^5$X, respectively, compared to conventional nonlinear materials, as in \figref{setup}(b,c). Nonlinear QWs have been applied to waveguide systems for frequency conversion \cite{vodopyanov1998phase} and THz laser generation \cite{vijayraghavan2013broadly}. Recently, Lee et al. \cite{lee2014giant,lee2016ultrathin,yu2019third} have demonstrated the coupling between QW and electromagnetic modes in metallic nanoresonators to realize nonlinear metasurfaces with high SHG and THG conversion efficiency. Yet there has been no accompanying theory to contextualize these results, and understand how close they might be to optimal. 

Conversely, for bulk crystals and organic materials, Thomas-Reiche-Kuhn (TRK) matrix-element sum rules represent strong constraints on linear \cite{shim2021fundamental} and nonlinear \cite{kuzyk2000physical,kuzyk2003erratum,kuzyk2000fundamental,kuzyk2003fundamental} susceptibilities. Kuzyk et al. have extensively studied the nonlinear response of organic molecules and derived upper bounds for off-resonant second-/third-order nonlinear responses based on three-/four-level ansatz and TRK sum rules \cite{kuzyk2000physical,kuzyk2003erratum,kuzyk2000fundamental,kuzyk2003fundamental}. Although these bounds significantly exceed the best nonlinear molecules by 30X for $\chi^{(2)}$ and 100X for $\chi^{(3)}$, creating the so-called "Kuzyk quantum gap", they scale correctly with physical properties of molecules (e.g., electron number, energy level spacing). Later theoretical models \cite{zhou2006pushing,zhou2007optimizing,lytel2013dressed,lytel2015phase,sullivan2016hybrid} further tightened these bounds. For linear materials, bounds on refractive index \cite{shim2021fundamental} have been found using dispersion-relation sum rules, effectively equivalent to TRK sum rules but somewhat more general (applying to any underlying interaction physics).

In this article, we establish theoretical bounds for on-resonant nonlinear susceptibilities of various frequency upconversion processes in QW systems, as in Fig. \ref{fig:setup}(b,c), based on TRK sum rules. We first connect the Kuzyk's oscillator strength bound \cite{kuzyk2006fundamental} with the on-resonant SHG $\chi^{(2)}$ bound, and show that many existing designed QWs can closely approach this bound. Next, we generalize the bound to the on-resonant SFG case, where current designs fall short of the bound by approximately 6X. We then present near optimum QW designs based on a proposed self-adjoint inverse design method for the 1D Schrodinger equation. Finally, we further utilize TRK sum rules to establish an on-resonant third-order sum-frequency generation ($\text{SFG}^{\text{(3)}}$) oscillator strength bound, and tighten the bound for the on-resonant THG case. We present near optimum QW designs using inverse design method that approach both $\text{SFG}^{\text{(3)}}$ and THG bounds. In the Conclusion, we summarize our work and discuss possible extensions of our framework to the bound for the nonresonant cases and the causal constructions of nonlinear susceptibility.

\section*{Bounds and designs for resonant $\chi^{(2)}$}
In this section we consider second-order nonlinear response. For resonant second-order nonlinearities, a single term dominates the susceptibility. Kuzyk has derived bounds for the oscillator strength of that term \cite{kuzyk2006fundamental}, while others have developed optimization techniques to identify QW designs with large nonlinearity. For second-harmonic generation, we show that established optimization techniques can closely approach the Kuzyk bounds. For sum-frequency generation, we show that a significant gap arises, but we present a general ``inverse-design'' framework for QWs that closely approaches (within 75\% to 95\%) the Kuzyk bound.

Considering the electronic contribution of the nonlinearity, a general expression for second-order nonlinear susceptibilities in quantum perturbation theory is a linear combination of generalized oscillators, comprising products of dipole matrix elements $x^{i}_{nm}$ (for $i$-polarized transitions between levels $n$ and $m$) divided by terms involving various $\omega_1$ and $\omega_2$, the optical frequencies, and $\omega_{nm}$ the transition frequencies between levels $n$ and $m$ (\citeasnoun{boyd2008nonlinear}):
\begin{equation} \label{eq:chi2}
\begin{split}
    \chi _{ijk}^{(2)}(\omega_1+\omega_2 ;\omega_1, \omega_2 ) = \frac{{N_e{e^3}}}{{{\varepsilon _0}{\hbar ^2}}} \mathcal{P}_{I}  \sum\limits_{n,m}{} [ \frac{{x_{gn}^i x_{nm}^j x_{mg}^k}}{{(\tilde{\omega} _{ng} - \omega_1-\omega_2 )({\tilde{\omega} _{mg}} - \omega_1 )}} +   \\
    \frac{x_{gn}^j  x_{nm}^k  x_{mg}^i }{(\tilde{\omega}^* _{ng} + \omega_2 )(\tilde{\omega}^* _{mg} + \omega_1 + \omega_2 )} +  \frac{x_{gn}^j x_{nm}^i x_{mg}^k }{(\tilde{\omega}^* _{ng} + \omega_2 )(\tilde{\omega} _{mg} - \omega_1 )} ] 
\end{split}
\end{equation}
where the variables $\tilde{\omega}_{nm}$ are complex-valued, adding an imaginary linewidth parameter to the transition frequency, $\tilde{\omega}_{nm}=\omega_{nm}-i \gamma_{nm}$, $\mathcal{P}_{I}$ is the intrinsic permutation operator, the transition dipole moments are $x_{nm}=\bra{n} x \ket{m}$ (where $x$ is the position three-vector, and superscripts isolate one component), and $N_e$ is the charge density. (The constant $e$ is the absolute value of the electron charge, $\hbar$ is the reduced Planck constant, and $\varepsilon_0$ is the free-space permittivity.) We assume the system is initially in a ground state $g$. 

The strongest second-order single-frequency (or narrowband) nonlinearities occur in the doubly resonant scenario when both terms in denominators of one of the oscillators in \eqref{chi2} are nearly zero. For example, in second-harmonic or sum-frequency generation, when $\omega_1 + \omega_2 \approx \Re(\tilde{\omega}_{ng})$ and $\omega_1 \approx \Re(\tilde{\omega}_{mg})$ for some $m$ and $n$, the first term in \eqref{chi2} will have a denominator approximately given by $-1/(\gamma_{ng}\gamma_{mg})$. For difference-frequency generation, $\omega_2$ can be considered negative, and one of the permutation terms would dominate the sum. Similarly for any second-order response; ultimately, one term in the summation dominates. We consider the diagonal components of nonlinear response with $i=j=k$. There is a simple argument for linear materials that the \emph{bounds} on anisotropic indices are largest ``on the diagonal''~\cite{shim2021fundamental} due to the eigenvalue nature of refractive indices and freedom of choice of principal axes, and this argument appears extensible---though we do not rigorously prove it---to the higher-order tensors of nonlinear optics through a generalization of the principal-axis idea~\cite{comon2008}. Dropping the implicit superscript and taking $i=j=k$, the assumption of doubly resonant response always yields a second-order susceptibility of the form
\begin{equation} \label{eq:maximum chi}
    \chi^{(2)}_{\text{res}} = \pm \frac{N_e e^3}{\epsilon_0 \hbar^2} \frac{x_{01}x_{12}x_{20}}{\gamma_{10}\gamma_{20}},
\end{equation}
where for simplicity we take the three levels involved in the double resonance to be the first three levels of the system ($0$, $1$, and 2), but the bounds below will not require this assumption (cf. SM). The sign ambiguity comes from whether the transition frequencies are conjugated or not in the resonant term. The linewidths $\gamma_{10}$, $\gamma_{20}$ are fixed by material and synthesis properties, leaving only the dipole matrix elements as the designable degrees of freedom. Bounds on the oscillator strength $x_{01} x_{12} x_{20}$ also lead directly to bounds on resonant $\chi^{(2)}$.

There are a variety of ``sum rules'' satisfied by dipole matrix elements \cite{wang1999generalization}, originating from commutator relations and asymptotic response limits. The most important ones for constraining response are the TRK sum rules \cite{thomas1925zahl,reiche1925zahl,kuhn1925gesamtstarke,bethe2012quantum,bethe2012quantum,jackiw2018intermediate}, which have been used for bounds for a variety of linear and nonlinear susceptibilities \cite{kuzyk2000physical,kuzyk2000fundamental,kuzyk2001quantum,kuzyk2003fundamental,kuzyk2003fundamental2,kuzyk2004doubly,kuzyk2006fundamental,shim2021fundamental}. They are derived by evaluating commutators $\bra{p}[[H,x],x]\ket{q}$ for quantum states $\ket{p}$ and $\ket{q}$. Leveraging the completeness of the quantum states, $\sum_n \ket{n}\bra{n}=I$, an infinite set of equations are obtained that encode constraints between dipole matrix elements and the energy spectra:
\begin{equation} \label{eq:TRK}
    \sum\nolimits_{n = 0}^\infty  {\left[ {{\omega _n} - {\textstyle{1 \over 2}}({\omega _p} + {\omega _q})} \right]x_{pn} x_{nq} = {\textstyle{{\hbar } \over {2m_e}}}{\delta _{pq}}} 
\end{equation}
where $m_e$ is the bare electron mass. For each pair $(p,q)$, the sum is over all states indexed by $n$. Interestingly, the TRK sum rules can also be interpreted as a consequence of causality-based sum rules applied to a dipole-moment-based linear susceptibility (cf. SM). (In a concurrent paper~\cite{Koutserimpas2024}, we identify new sum rules, derived either from causality or quantum-mechanics, for harmonic-generation processes. Whether they can further tighten bounds remains an open question.) To understand why TRK sum rules in particular are so useful for constraining response, consider the sum rule for $p=q=0$. In that case, \eqref{TRK} is a quadratic-form constraint with positive coefficients, forbidding arbitrarily large individual dipole moments. (That equation represents a finite-radius ellipsoid in the space of all dipole moments.) The additional constraints for other $(p,q)$ values can only further tighten possible response values.

Based on the TRK sum rules of \eqref{TRK}, Kuzyk has derived a bound on the $x_{01}x_{12}x_{20}$ oscillator strength \cite{kuzyk2006fundamental} by considering only the index pairs $(p,q)=(0,0)$ and $(p,q) = (1,1)$. These two pairs allow one to rewrite both $x_{20}$ and $x_{12}$ in terms of $x_{10}$ and 
many irrelevant matrix elements that only reduce oscillator strength. Once the total oscillator strength is written as a function only of $x_{10}$ (and the undesirable transitions), optimizing over $x_{10}$ leads to a bound on the magnitude of the oscillator strength that only depends on the TRK constant and the transition frequencies involved:
\begin{equation} \label{eq:bound-1}
\abs{x_{01}x_{12}x_{20}} \leq \frac{\sqrt[4]{3}  }{6} \left( \frac{\hbar}{m_e} \right)^{\frac{3}{2}} \frac{1}{\sqrt{\omega_{10}\omega_{21}\omega_{20}}} ,
\end{equation}
The bound is reached only in systems where all transitions to higher states vanish ($x_{ij}=0, \text{for } i=0,1,j>2$). (We include a simple derivation of \eqref{bound-1} in the SM.) Note that the scaling law for the oscillator strength is: $\abs{x_{01}x_{12}x_{20}}_{\rm max}\propto [(m_e\omega_{10})\cdot(m_e\omega_{21})\cdot(m_e\omega_{20})]^{-1/2}$. Generally, each dipole moment scales as $|x_{ij}|\sim (m_e\omega_{ij})^{-1/2}$, by taking $p=q=j$ in TRK sum rules \eqref{TRK}, as mentioned in \cite{kuzyk2013sum}. Alternatively, this can be derived from the scale invariance of Schodinger equation \cite{kuzyk2009bird}. As a result, the bound of \eqref{bound-1} indicates that a lower transition frequency produce a higher oscillator strength. Hence the bound for resonant nonlinear response is:
\begin{align}
    \left| \chi^{(2)}_{\text{res}} \right| \leq  \frac{\sqrt[4]{3}  }{6} \left( \frac{\hbar}{m_e} \right)^{\frac{3}{2}} \frac{N_e e^3}{\epsilon_0 \hbar^2 \gamma_{10}\gamma_{20}} \frac{1}{\sqrt{\omega_{10}\omega_{21}\omega_{20}}}  
\end{align}
This bound is valid for any resonant second-order nonlinear process that involves the transitions among lowest three states, including SHG, SFG, difference-frequency generation (DFG), and spontaneous parametric down-conversion (SPDC).

To consider the bound of \eqref{bound-1} in the context of quantum wells, one needs to replace the bare electron mass with the conduction band effective mass $m_e^*$. QWs are described by a one-dimensional Schrodinger equation with the effective mass taking the place of the bare mass, so that the usual TRK-sum-rule derivation still applies, now with $m_e^*$ replacing $m_e$ in the sum of \eqref{TRK}. This leads directly to $m_e^*$ replacing $m_e$ in the bound of \eqref{bound-1}. Given the resulting bound, the next step is to compare against various QW designs from across the literature. However, different designs with different material systems may have quite different effective masses; we find it useful to define a normalized oscillator strength that multiplies $|x_{01}x_{12}x_{20}|$ by $\left(m_e^*\right)^{3/2}$. This normalized oscillator strength is then subject to a bound that depends only on the constant $\hbar$ and the relevant frequencies of interest:
\begin{equation} \label{eq:bound-res-norm}
\abs{x_{01}x_{12}x_{20}} \left(m_e^*\right)^{3/2} \leq \frac{\sqrt[4]{3}\hbar^{3/2}}{6} \frac{1}{\sqrt{\omega_{10}\omega_{21}\omega_{20}}}.
\end{equation}
In the special case of SHG, defining $\omega = \omega_{10} = \omega_{21} = \omega_{20}/2$, the bound simplifies to:
\begin{equation} \label{eq:bound-shg}
\abs{x_{01}x_{12}x_{20}} \left(m_e^*\right)^{3/2} \leq \frac{\sqrt[4]{12}}{12} \left( \frac{\hbar}{\omega}\right)^{3/2}.
\end{equation}

\begin{figure*}[h]
    \begin{center}
    \includegraphics[width=1\linewidth]{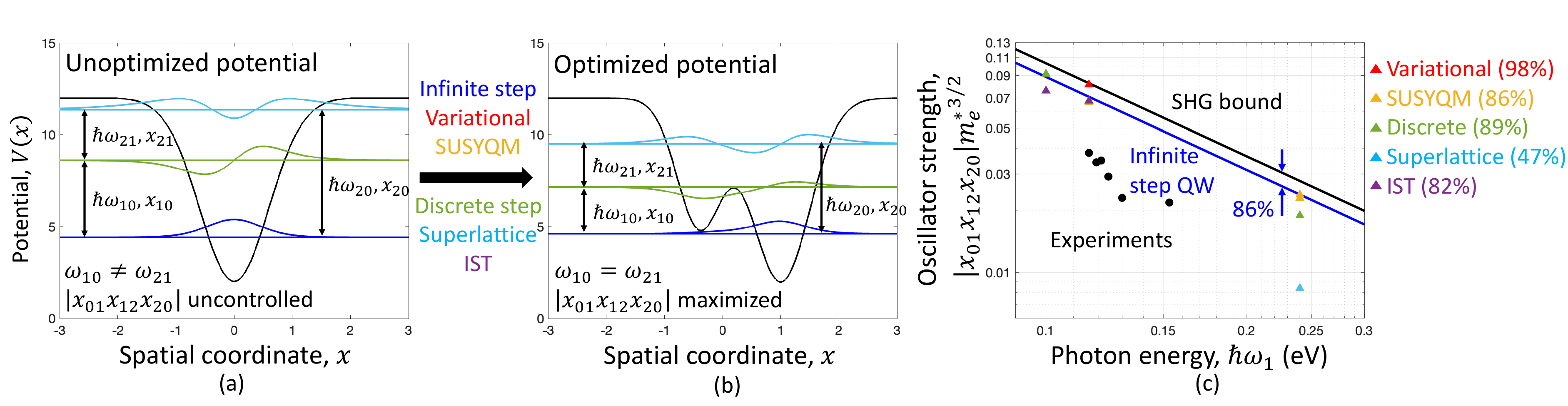}
    \end{center}
    \caption{Optimized QW structures can approach upper bounds to SHG. (a) A generic unoptimized potential has unequal energy-level spacings and uncontrolled oscillator strength $|x_{01}x_{12}x_{20}|$. (b) A variety of existing optimization methods can equalize the level spacing and maximize the oscillator strength. (c) Oscillator strength (normalized by effective mass) $\abs{x_{01}x_{12}x_{20}}\left(m_e^*\right)^{3/2}$ ($\SI{}{nm^3}\cdot m_e^{3/2}$) vs photon energy for the SHG bound, \eqref{bound-shg}, versus experimental measurements \cite{rosencher1991model,lee2014giant,lee2016ultrathin,kim2020spin,sarma2019broadband,rosencher1989second,capasso1994coupled} (black dots) and theoretical predictions (colored markers and blue line). The percentages in the parenthesis are the performances of best designs of a certain method. The single-step infinite QW design approach leverages scale invariance to produce near-optimal designs at any photon energy, achieving 86\% of the bound and optimal scaling. The variational approach (red triangle) even achieved 98\% of the bound for a specific photon energy $\SI{0.116}{eV}$.}
\label{fig:shg bound}
\end{figure*}

\Figref{shg bound} compares a variety of QW optimal designs against the bound of \eqref{bound-shg}. Over the past three decades, many QW design methods \cite{radovanovic2001two,indjin1998optimization,radovanovic2004quantum,indjin1998optimization,rosencher1991model,radovanovic1999resonant,radovanovic2001two,tomic1998quantum} produce potential profiles with both large oscillator strengths $x_{01}x_{12}x_{20}$ and equally spaced first three states, from the initial unoptimized QWs, as shown in \figref{shg bound}(a,b). We consider semi-analytical (``SUSYQM'' \cite{radovanovic2001two,indjin1998optimization,radovanovic2004quantum}, ``IST'' \cite{tomic1998quantum}), parameter-sweep (``Inifinite step'' \cite{rosencher1991model}, ``Discrete step'' \cite{indjin1998optimization,rosencher1991model}, and ``Superlattice'' \cite{radovanovic1999resonant}), and gradient-optimization techniques (``Variational'' \cite{radovanovic2001two}), and plot the parameters from the optimal designs in these references in \figref{shg bound}(c). All of the designs are for single photon frequencies (triangular markers) except that of the ``Infinite step'' approach (solid blue line), which exploits the infinite step for a simple reparametrization at any resonant frequency. Also included in black circular markers are experimentally measured oscillator strengths (appropriately normalized by effective masses). Plotted alongside the experimental data and theoretical designs is the bound of \eqref{bound-shg} (black solid line). The experimental measurements achieve up to 46\% of the bound; a variety of imperfections could be at fault, including non-parabolic bands \cite{sirtori1994nonparabolicity}. For the theoretical designs, one can see that the single data point for the ``Variational'' method reaches 98\% of the bound at $\hbar \omega = \SI{0.116}{eV} $. The ``Infinite step'' approach is within 20\% of the bound across all photon frequencies, suggesting that the bound of \eqref{bound-shg} is tight, or nearly so. Moreover, it shows that the scaling of the bound with $\omega^{-3/2}$ should be optimal.


\begin{figure*}[h]
\begin{center}
\includegraphics[width=1\linewidth]{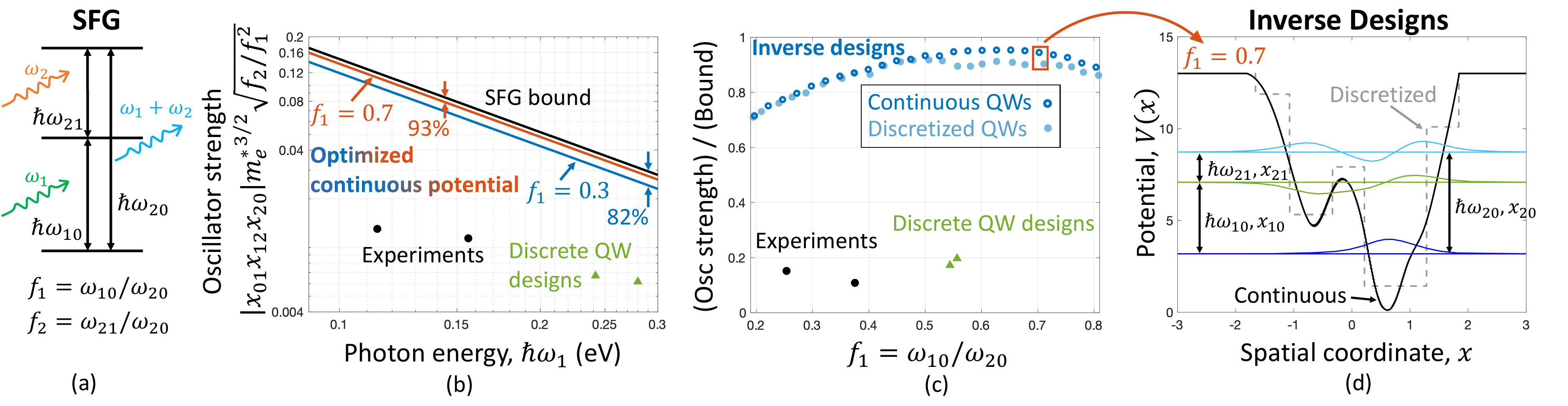}
\end{center}
\caption{\small Inversed designed QWs can approach the upper bounds to SFG. (a) Schematic representation of the spectral spacing $f-$factors. (b) Oscillator strength (normalized by effective mass and $f-$factors) $\abs{x_{01}x_{12}x_{20}} \left(m_e^*\right)^{3/2}\sqrt{{f_2}/{f_1^2}}$ ($\SI{}{nm^3}\cdot{m_e^{3/2}}$) vs photon energy for the SFG bound, \eqref{bound-sfg}, versus experimental measurements \cite{fejer1989observation,nefedkin2023overcoming} (black dots) and theoretical predictions \cite{yildirim2017second} (colored markers); also shown are the near-optimal continuous QWs with $f_1=0.7$ and $f_1=0.3$ by inverse design method (blue and orange lines), which extends across different photon frequencies due to scaling invariance. (c) Inverse designed and existing QW oscillator strength normalized by SFG bound, \eqref{bound-sfg}, vs level spacing $f-$factors. The bounds and optimal designs depicted here apply directly to DFG, etc. (d) Inverse designed near-optimal continuous QW with $f_1=0.7$ (black line, design/bound=93\%), and reduced discretized QW (gray dashed line, design/bound=91\%). The energy and spatial coordinate are in dimensionless units; to convert to specific units, consider if the spatial coordinate is in unit of $a$, then the energy should be in the unit of $\hbar^2/(m^*_ea^2)$.}
\label{fig:sfg bound}
\end{figure*}

Next we consider the case of resonant SFG with $\omega_{10}\neq \omega_{21}$. In this case, design approaches and experimental results from the literature are significantly farther from the bounds, but we describe an adjoint-based inverse design technique that leads to designs closely approaching the bounds. The bound of \eqref{bound-res-norm} applies directly to this case, but it is inconvenient to try to compare systems with different frequency spacings in a single plot. To simplify the comparisons, we define fractions $f_1$ and $f_2$ that represent the relative ratios of $\omega_{10}$ and $\omega_{21}$ to their sum, $\omega_{20}$: $f_1 = \omega_{10} / \omega_{20}$, $f_2 = \omega_{21} / \omega_{20}$. Note that $f_1 + f_2 = 1$, as in~\figref{sfg bound}(a). Then we can renormalize the oscillator strength by a factor $\sqrt{f_2/f_1^2}$, and replace $\omega_{10}$ with $\omega_1$ (the first field frequency, because the system is resonant), yielding the renormalized bound:
\begin{equation} \label{eq:bound-sfg}
\abs{x_{01}x_{12}x_{20}} \left(m_e^*\right)^{3/2} \sqrt{\frac{f_2}{f_1^2}} \leq  \frac{\sqrt[4]{3}}{6} \left( \frac{\hbar}{\omega_1} \right)^{3/2} 
\end{equation}
\Figref{sfg bound}(b) shows the SFG bound of \eqref{bound-sfg} in the solid black line. Of the many design techniques considered for SHG of~\figref{shg bound}, only the ``Discrete step'' paper \cite{yildirim2017second} considered the more general case of SFG, and its designs fall about 5X short of the bounds. Similarly, the best experimental measurements \cite{nefedkin2023overcoming} are all about 6X short of the bounds.

To test whether there exist QW designs that approach the bounds, we propose a gradient-based ``inverse design'' method. Adjoint-based inverse design has been successfully applied to many near-field photonics design problems~\cite{jensen2011topology,miller2012photonic,lu2013nanophotonic}, and its mathematical underpinnings have been used in control theory~\cite{pontryagin2018mathematical}, elasticity~\cite{bendsoe2013topology}, neural networks~\cite{werbos1994roots,lecun2015deep}, and quantum electronics~\cite{levi2010novel}, yet it appears to not have been applied to QW designs.

The central idea of the adjoint approach is to utilize ``reciprocity'' (transpose symmetry, or its generalization) so that gradients of an objective with respect to any number of geometrical degrees of freedom can be computed quickly and efficiently. In the case of quantum wells, the shape of the potential constitutes the degrees of freedom. For the $\chi^{(2)}$ response of this section, we define an objective function $F$ given by the oscillator strength, $F=x_{01} x_{12} x_{20}$ (we assume the lowest three states are bound states, so that $x_{01}, x_{12}, x_{20}$ are real values). 

A gradient is a variation of a function with respect to arbitrary variations of a parameter. For parameters that are functions, such as the potential $V(x)$, instead of an explicit gradient one often relates variations in the function, $\delta F$, to arbitrary variations in the parameter function, $\delta V(x)$. To connect the two in the case of QW design, we need the variations in the matrix elements $x_{ij}$ for any variation in the potential. (The variations in the function $F$ from variations in $x_{ij}$ then follow by the chain rule.) Variations in $x_{ij} = \mel{i}{x}{j}$ due to geometric perturbations are given by variations in the states $i$ and $j$:
\begin{align}
\delta x_{ij} = \mel{\delta i}{x}{j} + \mel{i}{x}{\delta j}.
\end{align}
The variations in the states, $\ket{\delta i}$ and $\ket{\delta j}$ are given by standard quantum perturbation theory \cite{sakurai2020modern}:
\begin{align}
    \ket{\delta i}&=\sum_{k\neq i}\frac{\bra{k}\delta V(x)\ket{i}}{E_i-E_k}\ket{k}.    
\end{align}
(Although not immediately obvious, this standard expression is the analog of the adjoint derivative discussed above \cite{miller2012photonic}, with the benefit that eigenvectors are ``self-adjoint'' in that a separate ``adjoint'' solution is not needed to compute the derivative.) Combining these variational expressions with the chain rule, and writing the real-space representations of states $\ket{i}$ as $\psi_i(x)$, we arrive at the objective-function variation:
\begin{align}
    \delta F = \int g(x) \delta V(x) \,{\rm d}x,
    \label{eq:df}
\end{align}
where
\begin{align}
    g(x) = x_{12} x_{20} \left(\sum_{k\neq 0} \frac{\psi_0^*(x) \psi_k(x) \mel{k}{x}{1}}{E_0 - E_k} + \sum_{k\neq 1} \frac{\psi_1^*(x) \psi_k(x) \mel{k}{x}{0}}{E_1 - E_k}\right) + \textrm{C.P.}
    \label{eq:gx}
\end{align}
where ``C.P.'' refers to identical terms with cyclic permutations of the indices, $(0,1) \rightarrow (1,2) \rightarrow (2,0) \rightarrow (0,1)$. \Eqref{df} now relates any variation in potential to the resulting change in the objective; the function $g(x)$ is referred to as a ``functional derivative''~\cite{courant2008methods}. For a finite-dimensional representation of $\delta V(x)$, e.g. $V(x) = \sum_i V_i \phi_i(x)$ for orthonormal $\phi_i(x)$, the variation would be written $\delta F = \sum_i g_i V_i$, where $g_i = \int \phi_i(x) g(x)$, and vector of $g_i$ values is a standard gradient.

Given the variational gradient of \eqreftwo{df}{gx}, one can then maximize the oscillator strength with any gradient-based optimization technique. During the optimization process it is important to keep fixed the level spacings $\omega_{10}$ and $\omega_{21}$, which may be dictated by the application of interest. There are many constrained optimization techniques that allow for such constraints, and they can be particularly efficient in this case because the gradients of the energies can be computed semi-analytically. By standard perturbation theory, the variation in a eigen-energy $E_i$ is given by~\cite{sakurai2020modern}:
\begin{align}
    \delta E_i =\bra{i} \delta V(x) \ket{i},
\end{align}
which can be used to ensure that the levels are properly spaced by the end of the optimization process.

We use the interior-point algorithm \cite{byrd2000trust,byrd1999interior} (implemented by MATLAB's ``fmincon'' function) to optimize the oscillator strength subject to fixed energy-level spacings. \Figref{sfg bound}(b) plots optimal design oscillator strengths (normalized as in \eqref{bound-sfg}) for two representative energy-level spacings given by $f_1 = 0.3$ and $f_1 = 0.7$. One can see that the solid lines for these designs quite closely approach the bounds, achieving 82\% and 93\%, respectively, across all fundamental photon frequencies considered. The potential is represented as a set of piecewise constant functions with coefficients that represent the degrees of freedom that are optimized. A high resolution (more than 200 points per potential) is used to approximate a continuous potential. \Figref{sfg bound}(c) compares the inverse-designed and previous-literature QW normalized oscillator strengths for varying $f_1$ values. (By scale invariance one need not fix the photon frequency as $f_1$ varies.) It also shows the discretized versions of the continuous designs. They are straightforward to discretize by taking the piecewise average of the continuous potential, and the small number of discrete-potential parameters can then be re-optimized with a gradient-based approach, where the gradient of each parameter is calculated in the same way as above. The 5\% degradation in performance from discretization is surprisingly small, and the energy levels can remain almost fixed at their values for the continuous counterpart, with $\sim2\%$ deviations. \Figref{sfg bound}(d) shows the continuous QW design (solid black) and its discretized counterpart (dashed gray), each of which achieve $> 90\%$ of the bound for $f_1 = 0.7$. One can see that the inverse design technique successfully approaches the bounds for a wide set of systems.

Resonant SFG in QWs has been shown to produce a higher nonlinear conversion efficiency compared to SHG \cite{nefedkin2023overcoming}, but the designed QWs reported in the literature only reach 15\% of the SFG bound for $f= 0.25, 0.72$. Leveraging our inverse design method, the optimized QWs can achieve an oscillator strength up to 4-5X higher, substantially enhancing the resonant SFG $\chi^{(2)}$ and nonlinear conversion efficiency. 

\begin{figure*}[h]
\begin{center}
\includegraphics[width=1\linewidth]{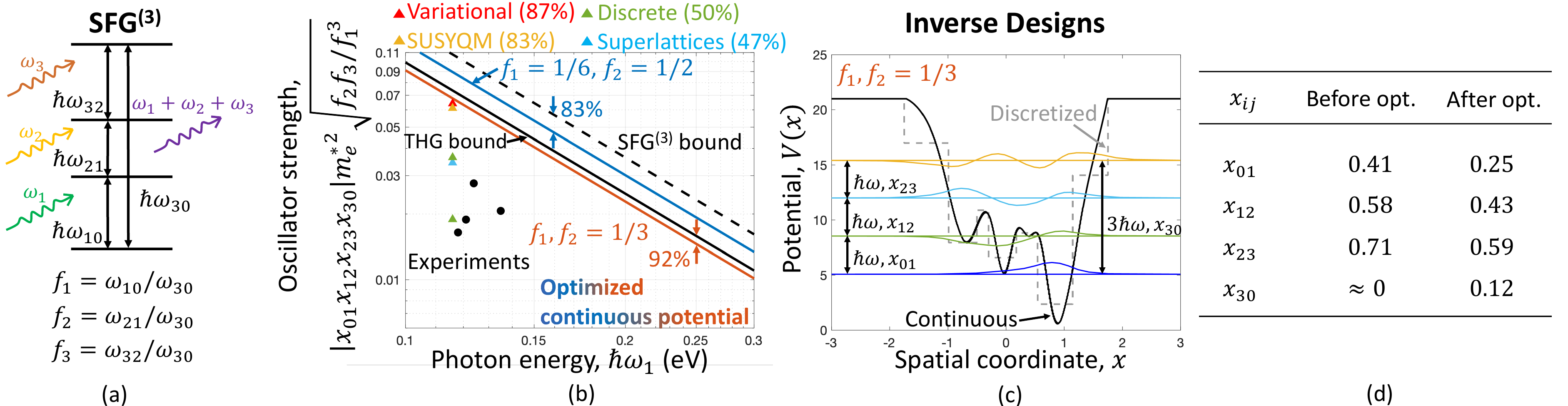}
\end{center}
\caption{\small Existing and inversed designed QWs can approach the upper bounds to $\text{SFG}^{\text{(3)}}$ and THG. (a) Schematic representation of the spectral spacing $f-$factors. (b) Oscillator strength (normalized by effective mass and $f-$factors) $\abs{x_{01}x_{12}x_{23}x_{30}} \left(m_e^*\right)^{2} \sqrt{{f_2f_3}/{f_1^3}}$ ($\SI{}{nm^4}\cdot{m_e^{2}}$) vs photon energy for $\text{SFG}^{\text{(3)}}$ (dashed line, \eqref{bound-sfg3}) and THG bounds (black solid line, \eqref{bound-thg}), versus experimental measurements \cite{yu2019third,kim2020spin,capasso1994coupled,zahedi2019design} (black dots) and theoretical predictions \cite{radovanovic2001global,indjin1998optimization,radovanovic1999resonant} (colored markers); also shown are the near-optimal continuous QWs with $\{f_1=1/6,f_2=1/2,f_3=1/3\}$ and $\{f_1=f_2=f_3=1/3\}$ by inverse design method (blue and orange lines). (c) Inverse designed near-optimal continuous QW for THG with $f_1=f_2=f_3=1/3$ (black line, design/bound=92\%), and reduced discretized QW (gray dashed line, design/bound=91\%). (d) Dipole matrix elements before and after inverse-design optimizations.} 
\label{fig:thg bound}
\end{figure*} 

\section*{Bounds and designs for resonant $\chi^{(3)}$}

In this section, we consider resonant third-order nonlinear process $\text{SFG}^{\text{(3)}}$ and THG. We apply TRK sum rules \eqref{TRK} to determine the resonant $\chi^{(3)}$ oscillator strength bounds, compare them with existing QWs, and present near-optimal QW designs based on the inverse design method described above. Unlike the SFG bound for $x_{01}x_{12}x_{20}$, the general $\text{SFG}^{\text{(3)}}$ bound for $x_{01}x_{12}x_{23}x_{30}$ can be further tightened with specific spectral configurations, e.g. equally spaced four lowest states used for THG. This is achieved by applying additional TRK sum rules with spectral constraints, and we present corresponding new bounds and near-optimal QW designs for THG.

Similar to the doubly resonant second-order nonlinear response of~\eqref{maximum chi}, the triply resonant response takes the form:
\begin{equation} \label{eq:maximum chi(3)}
    \chi^{(3)}_{\text{res}} = \pm \frac{N_e e^4}{\epsilon_0 \hbar^3} \frac{x_{01}x_{12}x_{23}x_{30}}{\gamma_{10}\gamma_{20}\gamma_{30}}
\end{equation}
where we assume the transition is among lowest four states for simplicity, and the oscillator strength $|x_{01}x_{12}x_{23}x_{30}|$ is the quantity to bound. For the SFG$^\text{(3)}$ case, we define fractions $f_1,f_2,f_3$ for simplicity: $f_1=\omega_{10}/\omega_{30}, f_2=\omega_{21}/\omega_{30}, f_3=\omega_{32}/\omega_{30}$, where $f_1+f_2+f_3=1$, as in~\figref{thg bound}(a). Leveraging the TRK sum rules of \eqref{TRK} with $(p,q)=(0,0),(1,1),(2,2)$, the resonant $\text{SFG}^{\text{(3)}}$ bound for $|x_{01}x_{12}x_{23}x_{30}|$ normalized by a factor of $\left(m_e^*\right)^2\sqrt{f_2f_3/f_1^3}$ is obtained (cf. SM):
\begin{equation} \label{eq:bound-sfg3}
    \abs{x_{01}x_{12}x_{23}x_{30}} \left(m_e^*\right)^{2} \sqrt{\frac{f_2f_3}{f_1^3}} \leq \frac{1}{4} \left( \frac{\hbar}{\omega_1} \right)^{2}.
\end{equation}
\Eqref{bound-sfg3} appears to not have been derived before; the closest expression is a bound for \emph{off-resonant} oscillator-strength terms for which different TRK sum rules comprise the key constraints \cite{kuzyk2000fundamental,kuzyk2004doubly}. The bound is reached only in systems where all transitions to higher states vanish ($x_{ij}=0, $\text{for }$ i=0,1,2, j>3$). It applies to resonant $\text{SFG}^{\text{(3)}}$ for any valid $\{f_1,f_2,f_3\}$ combinations. Unlike the SFG bound of~\eqref{bound-sfg}, it can be further tightened for the special case of resonant THG ($f_1=f_2=f_3=1/3$). The $\text{SFG}^{\text{(3)}}$ bound of~\eqref{bound-sfg3} can be tightened by including the additional TRK sum rules, $(p,q)=(0,1),(0,2),(1,2)$, which relates diagonal ($x_{11},x_{22}$) and off-diagonal ($x_{01},x_{02},x_{12}$) dipole matrix elements. Subsequently, an extra 0.67 prefactor is introduced to the $\text{SFG}^{\text{(3)}}$ bound of \eqref{bound-sfg3}, giving the THG bound (cf. SM):
\begin{equation} \label{eq:bound-thg}
    \abs{x_{01}x_{12}x_{23}x_{30}} \left(m_e^*\right)^{2} \sqrt{\frac{f_2f_3}{f_1^3}} \leq \frac{0.68}{4} \left( \frac{\hbar}{\omega_1} \right)^{2} 
\end{equation}
\Eqref{bound-thg} only applies for THG; for all other cases, the bound of \eqref{bound-sfg3} should be used.

\Figref{thg bound}(b) compares both bounds with various existing experimentally measured and theoretically designed QWs, where we only include data points for THG QWs due to a lack of data in designed $\text{SFG}^{\text{(3)}}$ ones. Among them, a single data point for ``Variational'' method \cite{radovanovic2001global} produces a theoretical design with the highest performance, achieving 87\% of THG bound at $\hbar \omega =\SI{0.116}{eV}$, while the best experimentally measured QW \cite{zahedi2019design} approaches the THG bound by 43\% at $\hbar\omega=0.124eV$, showing some room for future improvement. 

To find near-optimal QWs for THG and $\text{SFG}^{\text{(3)}}$, we use the inverse design method as discussed above. The designed continuous QWs for THG achieve the THG bound of~\eqref{bound-thg} by 92\%, as shown by red line in \figref{thg bound}(b), and its potential profile is shown in \figref{thg bound}(c). The performance of the best discretized QW only reduces performance by 1\%, while keeping the spectral configuration, again showing feasibility for experimental implementation. The oscillator strengths before and after the optimization are shown in \figref{thg bound}(d); the optimization clearly increases the value of $x_{30}$, which was practically zero before optimization. (The initial well is a simple, asymmetric quadratic-plus-cubic potential.) We also design near-optimal QWs for $\text{SFG}^{\text{(3)}}$ ($f_1=1/6,f_2=1/2,f_3=1/3$), which exceed the THG bound but are effectively constrained by the $\text{SFG}^{\text{(3)}}$ bound of~\eqref{bound-sfg3} within a factor of 20\%, as illustrated in Fig. \ref{fig:thg bound}(b). This confirms the validity and tightness of $\text{SFG}^{\text{(3)}}$ bound.

An interesting extension of the analysis above is to the case of a QW whose resonant transitions are not the lowest-lying levels, but rather higher-energy bound states. Typically in QWs it can be difficult to create large barrier heights that naturally yield many bound states, but wide wells are possible, and it is worth considering what susceptibility values would arise. The coefficients in the bounds derived above can change significantly as the resonant-level index increases. For example, if the resonant transitions occur between levels $n-2$ (or $n-3$) through $n$, the corresponding oscillator-strength bounds scale as $\sim n^{3/2}$ and $\sim n^2$ for $\chi^{(2)}$ and $\chi^{(3)}$, respectively. (Cf. the SM for a detailed derivation.) The physical intuition for this coefficient increase is that the relevant TRK sum rules of \eqref{TRK} have \emph{negative} contributions in the left-hand summation from levels below the resonant transitions (since $\omega_n$ is less than the average of $\omega_p$ and $\omega_q$), which implies the possibility for increasingly large transition elements via higher levels. 

\section*{Discussion and Outlook }
We have identified maximal resonant second- and third-order nonlinear optical response via electronic transitions. The TRK sum rules are relatively simple but strong constraints to oscillator strengths, and we show across a variety of cases that ``inverse-designed'' quantum wells can closely approach the bounds. Experimental demonstration of such performance could lead to 5X or larger increases beyond the state-of-the-art in sum-frequency generation, for example. Moreover, these bound and inverse design approaches can be applied to other resonant nonlinear process such as high-harmonic generation or single-photon down conversion (the reverse of SFG), potentially used in quantum cryptography, metrology and computing \cite{couteau2018spontaneous}, that can be possibly achieved via nonlinear optical metasurfaces \cite{solntsev2021metasurfaces,wang2022metasurfaces,santiago2022resonant}.

Looking forward, there are a number of avenues for further exploration. An important question is whether this approach, or another one, can yield similarly rigorous bounds to nonresonant nonlinear response. Though off resonance is weaker than resonant response, nonresonant nonlinearities can be strong over a wide bandwidth and have essentially no dissipative loss. Yet the application of TRK sum rules to general nonresonant response maximization is surprisingly tricky. For nonresonant response, susceptibility contributions from all levels should be included (not just the three or four involved in the resonant transitions). Yet when including more levels, the TRK sum rules allow for increasingly large low-level oscillator strengths as the number of high-energy levels increases. This leads to a ``many-level catastrophe,'' \cite{shafei2013paradox} as the corresponding bounds diverge as the number of allowed levels goes to infinity. (One cannot \emph{a priori} impose a restriction on the number of quantum bound and continuum states that contribute to nonlinear response.) This can be regularized with a statistical Monte Carlo approach \cite{lytel2017exact}, although this is due to the fact that such approaches do not find the global optima. Perhaps other sum rules or other as yet unidentified constraints are necessary to achieve a bound for nonresonant nonlinear response allowing for any number of levels.


Another interesting question is whether a nonlinear susceptibility of a form similar to \eqref{chi2} can be argued from more general principles than single-electron quantum mechanics. For example, a many-oscillator Drude-Lorentz representation of linear susceptibilities can be derived by causality through Kramers--Kronig relations~\cite{shim2021fundamental}, without any reference to specific quantum-mechanical assumptions. Many nonlinear susceptibility satisfy analogous Kramers--Kronig-like relations~\cite{lucarini2005kramers}, but the difficulty is in constraining the corresponding amplitudes. Quantum susceptibilities of the form of \eqref{chi2} represent nonlinear response in terms of the same dipole transition matrix elements that comprise linear susceptibilities, which then enables TRK-sum-rule constraints even for nonlinear response. It is not clear if there is an analog for causality-based representations of nonlinear susceptibilities.  


Finally, our results suggest that two materials-synthesis advancements could lead to significant increases in resonant nonlinearities. The first would be a dramatic reduction of the effective mass of the quantum well. The bounds of \eqref{bound-res-norm} and \eqref{bound-sfg3} predict maximum nonlinearities that scale as $1/\left(m_e^*\right)^{3/2}$ for $\chi^{(2)}$ and $1/\left(m_e^*\right)^2$ for $\chi^{(3)}$, meaning that 10X reductions in effective mass would lead to 32X and 100X increases in QW $\chi^{(2)}$ and $\chi^{(3)}$ susceptibilities, respectively. The second is a QW design whose resonant transitions occur between higher levels, rather than the lowest three/four. As discussed above, if one can access resonant transitions up to level $n$, the corresponding oscillator-strength bounds scale as $\sim n^{3/2}$ and $\sim n^2$ for $\chi^{(2)}$ and $\chi^{(3)}$, respectively. This offers another potential pathway to order-of-magnitude enhancements in optical nonlinearities. 

\vspace{50pt}
\noindent\textbf{Acknowledgments}: We thank Andrea Alu and Zeyu Kuang for helpful discussions regarding quantum well designs and bound formulations.

\noindent\textbf{Funding}: H.L. and O.D.M. were supported by Azimuth Corporation, Grant No.\ FA8650-16-D-5404, by Air Force Office of Scientific Research Grant No. FA9550-22-1-0393, and by the Simons Collaboration on Extreme Wave Phenomena Based on Symmetries (Award No.\ SFI-MPS-EWP-00008530-09). F.M. acknowledges support from the Air Force Office of Scientific Research with Grant No. FA9550-22-1-0204, through Dr. Arje. Nachman. T.\ T.\ K.\ is supported by the Swiss National Science Foundation (SNSF), Grant No.\ 203176, and the ``Stamatis G. Mantzavinos'' Postdoctoral Scholarship from the Bodossaki Foundation. 

\clearpage

\section*{Supplementary}

\subsection{Derivation of TRK sum rules with causality}
In this section, we derive a less general version of TRK sum rules~\eqref{TRK} in the main text based on the causality and dipole moment representation of linear susceptibility. We start with considering the Kramers-Kronig relations of the linear susceptibility, which relates its real part at one frequency to a principle value integral of imaginary part \cite{lucarini2005kramers}:
\begin{equation} \label{eq:KK}
    \Re \chi(\omega) = \frac{2}{\pi} \int_0^\infty \frac{\omega' \Im \chi(\omega')}{\omega'^2-\omega^2} \mathrm{d} \omega'
\end{equation}
Then the ``$f$-sum rule" is obtained by taking the asymptotic behavior of linear susceptibility to relate the integral of imaginary part of susceptibility to system's plasma frequency $\omega_p$ (\citeasnoun{king1976sum}):
\begin{equation} \label{eq:f-sum}
    \int_0^\infty {\omega \Im \chi(\omega)} \mathrm{d} \omega=\lim_{\omega \to \infty}\left(-\frac{\pi \omega^2}{2}\Re \chi(\omega)\right)=\frac{\pi e^2 N_e}{2\varepsilon_0 m_e} =\frac{\pi \omega_p^2}{2}
\end{equation}
\Eqref{f-sum} is valid for any linear, passive and causal material. Therefore, the dipole-moment-based linear susceptibility, derived by perturbation theory \cite{boyd2008nonlinear}, should satisfy this ``$f$-sum rule". We will then derive the TRK sum rules~\eqref{TRK} using this linear susceptibility.

The Eq. (3.2.23) in \citeasnoun{boyd2008nonlinear} gives $\chi(\omega)$ with the assumption that the ground state $g$ is the unperturbed state, which can be easily generated to the case where electrons occupy the $l-$th state as initial state. Assuming the polarization field is parallel to the electric field, then we can drop the polarization indices, giving:
\begin{equation} 
    \chi(\omega)=\frac{N_e e^2}{\varepsilon_0 \hbar} \sum_n \left( \frac{\abs{x_{ln}}^2}{\omega_{nl}-\omega-i\gamma_{nl}} + \frac{\abs{x_{ln}}^2}{\omega_{nl}+\omega+i\gamma_{nl}} \right)
\end{equation}
Combining it with \eqref{f-sum}, we can get:
\begin{equation} \label{eq:TRK diag}
    \boxed{\sum_n \omega_{nl}\abs{x_{ln}}^2=\frac{\hbar}{2m_e}}
\end{equation}
which is the TRK sum rules~\eqref{TRK} with $p=q=l$. Furthermore, if we assume the system's unperturbed state is the linear combination of the state $p$ and $q$ with probability $c_p$ and $c_q$, where $\abs{c_p}^2+\abs{c_q}^2=1$, then:
\begin{equation} \label{eq:initial}
    \psi^{(0)}(x,t)=c_p u_p(x)e^{-iE_p t/\hbar} + c_q u_q(x)e^{-iE_q t/\hbar}
\end{equation}
The first order perturbation wavefunction $\psi^{(1)}(x,t)$ is:
\begin{equation} \label{eq:initial}
    \psi^{(1)}(x,t)=\sum_n a_n^{(1)}(t)u_n(x)e^{-iE_nt/\hbar}
\end{equation}
where $a_n^{(1)}(t)$ is the probability amplitude of first order perturbation at state $n$ and time $t$, which is related to the external optical field at frequency $\omega$: $\tilde{E}(t)=E(\omega) e^{-i\omega t}$. So $a_n^{(1)}(t)$ is:
\begin{equation} \label{eq:an}
    a_n^{(1)}(t)=\frac{e}{\hbar} \left( c_p \cdot \frac{x_{np}\cdot E(\omega)}{\omega_{np}-\omega} e^{i(\omega_{np}-\omega)t} + c_q \cdot \frac{x_{nq}\cdot E(\omega)}{\omega_{nq}-\omega} e^{i(\omega_{nq}-\omega)t} \right)
\end{equation}
The first order contribution to polarization field $\langle\tilde{p}^{(1)} \rangle$ is:
\begin{equation} \label{eq:p}
    \langle\tilde{p}^{(1)} \rangle=\mel{\psi^{(0)}}{e\cdot x}{\psi^{(1)}} + \mel{\psi^{(1)}}{e\cdot x}{\psi^{(0)}}
\end{equation}
Combining Eq. (\ref{eq:initial},\ref{eq:an},\ref{eq:p}) gives the linear susceptibility:
\begin{equation} \label{eq:chi mix}
    \chi(\omega)=\frac{N_e e^2}{\varepsilon_0 \hbar} \left( \chi_{pp}(\omega) + \chi_{qq}(\omega)+\chi_{pq}(\omega+\omega_{pq}) + \chi_{qp} (\omega+\omega_{qp}) \right)
\end{equation}
where
\begin{align*} \label{eq:chi pq}
    &\chi_{pp}(\omega)=\abs{c_p}^2 \sum_n \left( \frac{\abs{x_{pn}}^2}{\omega_{np}-\omega-i\gamma_{np}} + \frac{\abs{x_{pn}}^2}{\omega_{np}+\omega+i\gamma_{np}} \right) \\
    &\chi_{qq}(\omega)=\abs{c_q}^2 \sum_n \left( \frac{\abs{x_{qn}}^2}{\omega_{nq}-\omega-i\gamma_{nq}} + \frac{\abs{x_{qn}}^2}{\omega_{nq}+\omega+i\gamma_{nq}} \right) \\
    &\chi_{pq}(\omega+\omega_{pq})=c_q^*c_p \sum_n \left( \frac{x_{qn}x_{np}}{\omega_{np}-\omega-i\gamma_{np}} + \frac{x_{qn}x_{np}}{\omega_{nq}+\omega+i\gamma_{nq}} \right) \\
    &\chi_{qp}(\omega+\omega_{qp})=c_p^*c_q \sum_n \left( \frac{x_{pn}x_{nq}}{\omega_{nq}-\omega-i\gamma_{nq}} + \frac{x_{pn}x_{nq}}{\omega_{np}+\omega+i\gamma_{np}} \right)
\end{align*}
Clearly, due to~\eqref{TRK diag}, $\frac{N_e e^2}{\varepsilon_0 \hbar} \left( \chi_{pp}(\omega) + \chi_{qq}(\omega) \right)$ satisfies the ``$f-$sum rule'', \eqref{f-sum}, leaving the other two terms to satisfy:
\begin{equation} \label{eq:f-sum0}
    \int_0^\infty {\omega \Im \left( \chi_{pq}(\omega+\omega_{pq}) + \chi_{qp}(\omega+\omega_{qp}) \right)} \mathrm{d} \omega  =\lim_{\omega \to \infty}\left(-\frac{\pi \omega^2}{2}\Re \left( \chi_{pq}(\omega+\omega_{pq}) + \chi_{qp}(\omega+\omega_{qp}) \right) \right)=0
\end{equation}
Taking the limit $\gamma \to 0$, and due to the arbitrariness of $c_p$ and $c_q$, we have:
\begin{equation} \label{eq:TRK off-diag}
    \boxed{\sum_n (\omega_{np}+\omega_{nq})(x_{pn}x_{nq}+x_{qn}x_{np})=0,}
\end{equation}
which is the sum of terms $(p,q)$ and $(q,p)$ of the TRK sum rules of \eqref{TRK}. In this way, we can see that the TRK sum rules can be interpreted based on the causality and dipole moment representation of the linear susceptibility.

\subsection{Derivation of bound of oscillator strength $\abs{x_{01}x_{12}x_{20}}$}
In this section, we derive the bound of $\abs{x_{01}x_{12}x_{20}}$~\eqref{bound-1} in a simpler way than Kuzyk's method \cite{kuzyk2006fundamental}. Consider the TRK sum rule~\eqref{TRK} with $(p,q)=(0,0)$ and $(p,q)=(1,1)$:
\begin{equation}
    \begin{aligned} \label{eq:TRK 0011}
    &\omega_{10}\abs{x_{10}}^2 + \omega_{20}\abs{x_{20}}^2 + \sum_{n\geq3}\omega_{n0}\abs{x_{n0}}^2 = \frac{\hbar}{2m_e} \\
    &-\omega_{10}\abs{x_{10}}^2 + \omega_{21}\abs{x_{21}}^2 + \sum_{n\geq3}\omega_{n1}\abs{x_{n1}}^2= \frac{\hbar}{2m_e}
\end{aligned}
\end{equation}
The terms $\sum_{n\geq3}\omega_{n0}\abs{x_{n0}}^2$ and $\sum_{n\geq3}\omega_{n1}\abs{x_{n1}}^2$ are non-negative. Then $x_{20}$ and $x_{21}$ can be represented by $x_{10}$ and other irrelevant matrix elements:
\begin{equation}
    \begin{aligned} 
    &\abs{x_{20}} ={\frac{1}{\sqrt{\omega_{20}}}}\sqrt{\frac{\hbar}{2m_e} -\omega_{10}\abs{x_{10}}^2- \sum_{n\geq3}\omega_{n0}\abs{x_{n0}}^2 }  \\
    &\abs{x_{21}} ={\frac{1}{\sqrt{\omega_{21}}}}\sqrt{\frac{\hbar}{2m_e} + \omega_{10}\abs{x_{10}}^2- \sum_{n\geq3}\omega_{n1}\abs{x_{n1}}^2 }
\end{aligned} \label{eq:TRK x20x21}
\end{equation}
When $\sum_{n\geq3}\omega_{n0}\abs{x_{n0}}^2=\sum_{n\geq3}\omega_{n1}\abs{x_{n1}}^2=0$, and $\abs{x_{10}}=\frac{1}{\sqrt[4]{3}}\frac{\hbar}{2m_e \omega_{10}}$, the maximum oscillator strength $\abs{x_{01}x_{12}x_{20}}$ is obtained:
\begin{equation} 
\boxed{\abs{x_{01}x_{12}x_{20}} \leq \frac{\sqrt[4]{3}}{6} \left( \frac{\hbar}{m_e} \right)^{\frac{3}{2}} \frac{1}{\sqrt{\omega_{10}\omega_{21}\omega_{20}}}.}
\end{equation}
as in \eqref{bound-1}. Note that the optimal condition shows that the maximum oscillator strength is reached only when all dipole transition moments from $0-$/$1-$state to higher level vanish, indicating a simplified 3-level state system.

\subsection{Oscillator strength bound for other three-level combinations}
In the main text, we consider the oscillator strength bound with transitions among lowest three levels; here we formulate bounds for transitions among arbitrary three-level combinations, where the bound increases as the level index of the first and second transition state increases.

\subsubsection{Consecutive three-level combinations}
To start with, we first consider the transition among level 1, 2, and 3, denoted as $d_3=x_{12}x_{23}x_{31}$, and take the TRK sum rules with $(p,q)=(0,0),(1,1),(2,2)$:
\begin{equation}
    \begin{aligned}
    &\omega_{10}\abs{x_{10}}^2 + \omega_{20}\abs{x_{20}}^2 + \omega_{30}\abs{x_{30}}^2 + \sum_{n\geq4}\omega_{n0}\abs{x_{n0}}^2 = \frac{\hbar}{2m_e} \\
    &-\omega_{10}\abs{x_{10}}^2 + \omega_{21}\abs{x_{21}}^2 + \omega_{31}\abs{x_{31}}^2+ \sum_{n\geq4}\omega_{n1}\abs{x_{n1}}^2= \frac{\hbar}{2m_e} \\
    &-\omega_{20}\abs{x_{20}}^2 - \omega_{21}\abs{x_{21}}^2 + \omega_{32}\abs{x_{32}}^2 + \sum_{n\geq4}\omega_{n2}\abs{x_{n2}}^2= \frac{\hbar}{2m_e} 
\end{aligned}
\end{equation}
By substituting $x_{31}$ and $x_{32}$ with $x_{10}$ and $x_{21}$, we have:
\begin{align*}
    &\abs{x_{31}} ={\frac{1}{\sqrt{\omega_{31}}}}\sqrt{\frac{\hbar}{2m_e} + \omega_{10}\abs{x_{10}}^2- \omega_{21}\abs{x_{21}}^2 -\sum_{n\geq4}\omega_{n1}\abs{x_{n1}}^2 }  \\
    &\abs{x_{32}} ={\frac{1}{\sqrt{\omega_{32}}}}\sqrt{\frac{\hbar}{2m_e} + \left( \frac{\hbar}{2m_e}-\omega_{10}\abs{x_{10}}^2 -\sum_{n\geq3}\omega_{n0}\abs{x_{n0}}^2 \right) + \omega_{21}\abs{x_{21}}^2 -  \sum_{n\geq4}\omega_{n2}\abs{x_{n2}}^2 }
\end{align*}
The maximum $\abs{d_3}$ is obtained when:
\begin{align*}
    &\sum_{n\geq3}\omega_{n0}\abs{x_{n0}}^2=0 \\
    &\sum_{n\geq4}\omega_{n1}\abs{x_{n1}}^2=0 \\
    &\sum_{n\geq4}\omega_{n2}\abs{x_{n2}}^2=0 \\
    &\omega_{10}\abs{x_{10}}^2=\frac{\hbar}{2m_e} \\
    &\omega_{21}\abs{x_{21}}^2=\frac{1+\sqrt{7}}{3}\cdot \frac{\hbar}{2m_e}
\end{align*}
which result the bound:
\begin{equation} 
\abs{d_3}=\abs{x_{12}x_{23}x_{31}} \leq \sqrt{\frac{20+14\sqrt{7}}{27}} \left( \frac{\hbar}{2m_e} \right)^{\frac{3}{2}} \frac{1}{\sqrt{\omega_{31}\omega_{32}\omega_{21}}}.
\end{equation}
Similar as bound for $x_{01}x_{12}x_{20}$, \eqref{bound-1}, this bound is obtained when all dipole transition moments from 1-/2-state to higher level vanish; but it also requires that the dipole moment $x_{01}$ reaches it maximum: $\omega_{10}\abs{x_{01}}^2=\hbar/(2m_e)$. Due to this, the bound of $\abs{d_3}$ is 2.34X higher than that of $\abs{d_2}=\abs{x_{01}x_{12}x_{20}}$.

Furthermore, the bound can be generalized to the transitions among level $n-2,n-1$ and $n$, which is denoted as $d_n = x_{n-2,n-1}x_{n-1,n}x_{n,n-2}$. Taking the TRK sum rules with $(p,q)=(0,0),(1,1),\ldots,(n-1,n-1)$, and properly tuning the dipole moments other lower levels, we can have:
\begin{align*}
    \abs{d_n}  
    \leq &\sqrt{\left( \omega_{n-1,n-2}\abs{x_{n-1,n-2}}^2\right) \left( \frac{(n-1)\hbar}{2m_e}-\omega_{n-1,n-2}\abs{x_{n-1,n-2}}^2 \right) \left( \frac{\hbar}{2m_e}+\omega_{n-1,n-2}\abs{x_{n-1,n-2}}^2 \right) }  \\ 
    & \times \frac{1}{\sqrt{\omega_{n-1,n-2}\omega_{n,n-1}\omega_{n,n-2}}}  
\end{align*}
The bound is obtained when $\omega_{n-1,n-2}\abs{x_{n-1,n-2}}^2=\frac{1}{3}\left( \sqrt{n^2-n+1}+n-2 \right) {\frac{\hbar}{2m_e}}$, then:
\begin{align} \label{eq:consecutive 3}
    \abs{d_n}  
    \leq \frac{\sqrt{2n^3-3n^2+2(n^2-n+1)^{3/2}-3n+2}}{\sqrt{27}} \left( \frac{\hbar}{2m_e} \right)^{3/2} \frac{1}{\sqrt{\omega_{n-1,n-2}\omega_{n,n-1}\omega_{n,n-2}}}.  
\end{align}
The key result of this section is a scaling law for $|d_n|$: \mybox{$|d_n|\sim n^{3/2}$}. Existing QW designs focus on the transition of $d_1=x_{01}x_{12}x_{20}$, but this oscillator strength can be further enhanced by properly doping and designing QWs so that the transitions among higher levels are possible, thus $d_n$ can be used to produce a higher resonant nonlinearity. For example, if the QW is designed to produce a resonant response among level 9,10,11, then it can yield at most 21X higher oscillator strength than that of the lowest three states.

\subsubsection{Non-consecutive three-level combinations}
In this case, we consider the transition among three non-consecutive levels. For example, $x_{01}x_{13}x_{30}$, its bound is obtained by taking TRK sum rules with $(p,q)=(0,0),(1,1)$:
\begin{equation}
    \begin{aligned}
    &\omega_{10}\abs{x_{10}}^2 + \omega_{20}\abs{x_{20}}^2 + \omega_{30}\abs{x_{30}}^2 + \sum_{n\geq4}\omega_{n0}\abs{x_{n0}}^2 = \frac{\hbar}{2m_e} \\
    &-\omega_{10}\abs{x_{10}}^2 + \omega_{21}\abs{x_{21}}^2 + \omega_{31}\abs{x_{31}}^2+ \sum_{n\geq4}\omega_{n1}\abs{x_{n1}}^2= \frac{\hbar}{2m_e} 
\end{aligned}
\end{equation}
Again, $x_{30}$ and $x_{31}$ can be represented as $x_{10}$ which is similar as $x_{20}$ and $x_{21}$ in~\eqref{TRK x20x21}. The bound is the same as lowest three-level case, as in~\eqref{bound-1}, except that the optimal condition becomes $\omega_{20}\abs{x_{20}}^2 + \sum_{n\geq4}\omega_{n0}\abs{x_{n0}}^2=0$ and $\omega_{21}\abs{x_{21}}^2 + \sum_{n\geq4}\omega_{n1}\abs{x_{n1}}^2=0$. This means that the selection of the highest state will not affect the bound. Generally, the bound of oscillator strength $x_{l,l+1}x_{l+1,m}x_{m,l}$ ($l+1<m$) will require TRK sum rules with $(p,q)=(0,0),(1,1),\ldots,(l,l),(l+1,l+1)$, and the bound will equal to the bound of $x_{l,l+1}x_{l+1,l+2}x_{l+2,l}$.

On the other hand, if we consider the transition among state 0,2,3, or oscillator strength $x_{02}x_{23}x_{30}$, we should apply TRK sum rules with $(p,q)=(0,0),(1,1),(2,2)$:
\begin{equation}
    \begin{aligned} \label{eq:TRK 001122}
    &\omega_{10}\abs{x_{10}}^2 + \omega_{20}\abs{x_{20}}^2 + \omega_{30}\abs{x_{30}}^2 + \sum_{n\geq4}\omega_{n0}\abs{x_{n0}}^2 = \frac{\hbar}{2m_e} \\
    &-\omega_{10}\abs{x_{10}}^2 + \omega_{21}\abs{x_{21}}^2 + \omega_{31}\abs{x_{31}}^2 + \sum_{n\geq4}\omega_{n1}\abs{x_{n1}}^2= \frac{\hbar}{2m_e} \\
    &-\omega_{20}\abs{x_{20}}^2 - \omega_{21}\abs{x_{21}}^2 + \omega_{32}\abs{x_{32}}^2 + \sum_{n\geq4}\omega_{n2}\abs{x_{n2}}^2= \frac{\hbar}{2m_e}.
    \end{aligned}
\end{equation}
By substituting $x_{23}$, $x_{30}$ with $x_{02}$, $x_{01}$, we have:
\begin{align*}
    &\abs{x_{03}} ={\frac{1}{\sqrt{\omega_{31}}}}\sqrt{\frac{\hbar}{2m_e} - \omega_{10}\abs{x_{10}}^2- \omega_{20}\abs{x_{20}}^2 -\sum_{n\geq4}\omega_{n0}\abs{x_{n0}}^2 }  \\
    &\abs{x_{32}} ={\frac{1}{\sqrt{\omega_{32}}}}\sqrt{\frac{\hbar}{2m_e} + \left( \frac{\hbar}{2m_e}+\omega_{10}\abs{x_{10}}^2 -\sum_{n\geq3}\omega_{n1}\abs{x_{n1}}^2 \right) + \omega_{20}\abs{x_{20}}^2 -  \sum_{n\geq4}\omega_{n2}\abs{x_{n2}}^2 }
\end{align*}
The maximum $\abs{x_{02}x_{23}x_{30}}$ is obtained when:
\begin{align*}
    &\sum_{n\geq4}\omega_{n0}\abs{x_{n0}}^2=0 \\
    &\sum_{n\geq3}\omega_{n1}\abs{x_{n1}}^2=0 \\
    &\sum_{n\geq4}\omega_{n2}\abs{x_{n2}}^2=0 \\
    &\omega_{10}\abs{x_{10}}^2=0 \\
    &\omega_{20}\abs{x_{20}}^2=\frac{\sqrt{7}-1}{3}\cdot \frac{\hbar}{2m_e}
\end{align*}
which result the bound:
\begin{equation} 
\abs{x_{02}x_{23}x_{30}} \leq \sqrt{\frac{14\sqrt{7}-20}{27}} \left( \frac{\hbar}{2m_e} \right)^{\frac{3}{2}} \frac{1}{\sqrt{\omega_{31}\omega_{32}\omega_{21}}}.
\end{equation}
In this case, the bound requires the dipole moment $x_{01}$ vanishes, and this bound is 1.28X higher than that of $\abs{x_{01}x_{12}x_{20}}$. Again, this can be generalized to oscillator strength $x_{0,l}x_{l,m}x_{m,0}$ ($l<m$), i.e. the transitions among state $0,l$ and $m$. Taking the TRK sum rules with $(p,q)=(0,0),(1,1),\ldots,(l,l)$, and properly tuning the dipole moments among irrelevant levels, we have:
\begin{align*}
    \abs{x_{0,l}x_{l,m}x_{m,0}}  
    \leq &\sqrt{\left( \omega_{l,0}\abs{x_{0,l}}^2\right) \left( \frac{l\hbar}{2m_e}+\omega_{l,0}\abs{x_{0,l}}^2 \right) \left( \frac{\hbar}{2m_e}-\omega_{l,0}\abs{x_{0,l}}^2 \right) }  \\ 
    & \times \frac{1}{\sqrt{\omega_{l,0}\omega_{m,l}\omega_{m,0}}}  
\end{align*}
The bound is obtained when $\omega_{l,0}\abs{x_{0,l}}^2=\frac{1}{3}\left( \sqrt{l^2+l+1}-l+1 \right) {\frac{\hbar}{2m_e}}$, then:
\begin{align} \label{eq:non-consecutive 3}
    \abs{x_{0,l}x_{l,m}x_{m,0}}  
    \leq \frac{\sqrt{-2l^3-3l^2+2(l^2+l+1)^{3/2}+3l+2}}{\sqrt{27}} \left( \frac{\hbar}{2m_e} \right)^{3/2} \frac{1}{\sqrt{\omega_{l,0}\omega_{m,l}\omega_{m,0}}}  
\end{align} 
which has a similar form as the bound of \eqref{consecutive 3}, but a sign difference arise in the cubic term of the numerator, resulting this bound scaling as $\sim l^{1/2}$. 

More generally, for the oscillator strength of arbitrary levels, $x_{lm}x_{mn}x_{nl}$ ($l<m<n$), the bound will require the TRK sum rules with $(p,q)=(0,0),(1,1),\ldots,(l,l),\ldots,(m,m)$. The bound is independent of where is the $n-$th level. Then by properly tuning the dipole moments among irrelevant levels, we have:
\begin{align*}
    \abs{x_{lm}x_{mn}x_{nl}}  
    \leq &\sqrt{\left( \omega_{ml}\abs{x_{lm}}^2\right) \left( \frac{(m-l)\hbar}{2m_e}+\omega_{ml}\abs{x_{lm}}^2 \right) \left( \frac{(1+l)\hbar}{2m_e}-\omega_{ml}\abs{x_{lm}}^2 \right) }  \\ 
    & \times \frac{1}{\sqrt{\omega_{ml}\omega_{nm}\omega_{nl}}}  
\end{align*}
Define $a=l+1,b=m-l$, when $\omega_{ml}\abs{x_{lm}}^2$ satisfies:
\begin{align*}
    \omega_{ml}\abs{x_{lm}}^2=\frac{ \sqrt{a^2+ab+b^2}+a-b}{3} {\frac{\hbar}{2m_e}},
\end{align*}
the bound of oscillator strength $\abs{x_{lm}x_{mn}x_{nl}}$ is given by:
\begin{equation} \label{eq:arb 3 level}
    \begin{aligned}
    \abs{x_{lm}x_{mn}x_{nl}}    
    \leq &\frac{\sqrt{2a^3-2b^3+3a^2b-3ab^2+2(a^2+ab+b^2)^{3/2}}}{\sqrt{27}} 
    \left( \frac{\hbar}{2m_e} \right)^{3/2} \frac{1}{\sqrt{\omega_{ml}\omega_{nm}\omega_{nl}}}  
\end{aligned}
\end{equation}
The bound scales as $\sim(l+1)^{3/2}$ and $\sim(m-l)^{1/2}$, which conforms the above bounds of~\eqreftwo{consecutive 3}{non-consecutive 3}. From this, we can see that for the QW system with finite number of bound states, if one wants a higher oscillator strength, it is better to increase the level index of the first state than that of the second one.

\subsection{Derivation of bound of oscillator strength $\abs{x_{01}x_{12}x_{23}x_{30}}$}

In this section, we derive the bound of $\abs{x_{01}x_{12}x_{23}x_{30}}$ with TRK sum rules of~\eqref{TRK} for the general third-order sum-frequency generation (SFG$^{\text{(3)}}$), and we show the bound can be further tightened for the case of third-harmonic generation (THG). Consider the TRK sum rules of~\eqref{TRK} with $(p,q)=(0,0)$, $(p,q)=(1,1)$ and $(p,q)=(2,2)$:
\begin{equation}
    \begin{aligned} \label{eq:TRK 001122}
    &\omega_{10}\abs{x_{10}}^2 + \omega_{20}\abs{x_{20}}^2 + \omega_{30}\abs{x_{30}}^2 + \sum_{n\geq4}\omega_{n0}\abs{x_{n0}}^2 = \frac{\hbar}{2m_e} \\
    &-\omega_{10}\abs{x_{10}}^2 + \omega_{21}\abs{x_{21}}^2 + \omega_{31}\abs{x_{31}}^2 + \sum_{n\geq4}\omega_{n1}\abs{x_{n1}}^2= \frac{\hbar}{2m_e} \\
    &-\omega_{20}\abs{x_{20}}^2 - \omega_{21}\abs{x_{21}}^2 + \omega_{32}\abs{x_{32}}^2 + \sum_{n\geq4}\omega_{n2}\abs{x_{n2}}^2= \frac{\hbar}{2m_e}.
    \end{aligned}
\end{equation}
The terms The terms $\sum_{n\geq4}\omega_{n0}\abs{x_{n0}}^2$, $\sum_{n\geq4}\omega_{n1}\abs{x_{n1}}^2$, and $\sum_{n\geq4}\omega_{n2}\abs{x_{n2}}^2$ are non-negative. The dipole matrix moments $x_{12},x_{23},x_{30}$ can be represented by $x_{01}$ and other irrelevant matrix elements:
\begin{equation*}
    \begin{aligned}
    &\abs{x_{12}} ={\frac{1}{\sqrt{\omega_{21}}}}\sqrt{\frac{\hbar}{2m_e} + \omega_{10}\abs{x_{10}}^2 - \omega_{31}\abs{x_{31}}^2 - \sum_{n\geq4}\omega_{n1}\abs{x_{n1}}^2 }  \\
    &\abs{x_{23}} ={\frac{1}{\sqrt{\omega_{32}}}}\sqrt{\frac{\hbar}{m_e} + \omega_{10}\abs{x_{10}}^2 + \omega_{20}\abs{x_{20}}^2 - \omega_{31}\abs{x_{31}}^2 - \sum_{n\geq4}\omega_{n1}\abs{x_{n1}}^2 - \sum_{n\geq4}\omega_{n2}\abs{x_{n2}}^2 } \\
    &\abs{x_{30}} ={\frac{1}{\sqrt{\omega_{30}}}}\sqrt{\frac{\hbar}{2m_e} - \omega_{10}\abs{x_{10}}^2 - \omega_{20}\abs{x_{20}}^2 - \sum_{n\geq4}\omega_{n0}\abs{x_{n0}}^2 }.
\end{aligned}
\end{equation*}
When $\sum_{n\geq4}\omega_{n0}\abs{x_{n0}}^2=\sum_{n\geq4}\omega_{n1}\abs{x_{n1}}^2=\sum_{n\geq4}\omega_{n2}\abs{x_{n2}}^2=0$, $\omega_{20}\abs{x_{20}}^2=\omega_{31}\abs{x_{31}}^2=0$, and $\omega_{10}\abs{x_{01}}^2={\frac{(\sqrt{5}-1)\hbar}{4m_e }}$, the maximum oscillator strength $\abs{x_{01}x_{12}x_{23}x_{30}}$ is obtained:
\begin{equation} \label{eq:4l SFG}
\boxed{\abs{x_{01}x_{12}x_{23}x_{30}} \leq \frac{1}{4} \left( \frac{\hbar}{m_e} \right)^{2} \frac{1}{\sqrt{\omega_{10}\omega_{21}\omega_{32}\omega_{30}}},}
\end{equation}
which is the~\eqref{bound-sfg3} when replace $\omega_{21},\omega_{32},\omega_{30}$ with $\omega_{10}$ and $f-$factors. This bound also implies that the existence of irrelevant dipole matrix elements will only reduce the oscillator strength, and the bound is reached only in a 4-level system where all dipole transition moments from 0-/1-/2-state to higher state vanish.

Next, we consider the case of resonant THG, where $\omega_{10}=\omega_{21}=\omega_{32}$, and apply additional TRK sum rules besides the above-mentioned ones to tighten the bound. Assuming, for simplicity, the lowest 4 states are bound state, so that we take the corresponding dipole moments as real values. For TRK sum rules of~\eqref{TRK} with $(p,q)=(0,1),(0,2),(1,2)$:
\begin{equation}
    \begin{aligned} \label{eq:TRK 010212}
    &\omega_{10}x_{10}(x_{11}-x_{00}) + (\omega_{21}+\omega_{20})x_{02}x_{21} + (\omega_{31}+\omega_{30})x_{03}x_{31} = 0 \\
    &\omega_{20}x_{20}(x_{22}-x_{00}) + (\omega_{10}-\omega_{21})x_{01}x_{12} + (\omega_{30}+\omega_{32})x_{03}x_{32} = 0 \\
    &-(\omega_{10}+\omega_{20})x_{10}x_{02} + \omega_{21}(x_{22}-x_{11})x_{12} + (\omega_{31}+\omega_{32})x_{13}x_{32} = 0.
\end{aligned}
\end{equation}
Since the bound is achieved only in 4-level systems, we ignore the dipole moments terms with transitions to higher levels. These sum rules reveal the interdependence of off-diagonal dipole matrix elements ($x_{ij}$) through diagonal ones ($x_{ii}-x_{00}$), so that $x_{20}$ and $x_{31}$ cannot be 0 simultaneously, then the bound of~\eqref{4l SFG} can be tightened. Therefore, we can cancel the diagonal terms ($x_{11}-x_{00}$ and $x_{22}-x_{00}$) and have:
\begin{align}
    -2x_{01}x_{12}x_{23}x_{30}-3x_{01}^2x_{02}^2 + 3x_{02}^2x_{12}^2 + 5x_{03}x_{31}x_{12}x_{20} + 3x_{13}x_{32}x_{10}x_{20}=0,
\end{align}
Together with~\eqref{TRK 001122}, we have 4 equality constraints and 6 variables. Since its hard to solve the optimization problem analytically, we numerically find the global maximum of oscillator strength. We randomly generate 1 million initial points, and use the interior-point algorithm (implemented by MATLAB's ``fmincon'' function) as a local optimizer to find corresponding local optimal. We find around 93\% of initial guesses converge to a same local maximum (except for a sign ambiguity of $x_{ij}$), which is higher than any other local maximum. Then we believe the global maximum is reached, which introduces an extra 0.68 prefactor to the original bound of \eqref{4l SFG}, yielding:
\begin{equation} \label{eq:4l THG}
\boxed{\abs{x_{01}x_{12}x_{23}x_{30}} \leq \frac{0.68}{4} \left( \frac{\hbar}{m_e} \right)^{2} \frac{1}{\sqrt{\omega_{10}\omega_{21}\omega_{32}\omega_{30}}},}
\end{equation}
which is the~\eqref{bound-thg} in the main text when replace $\omega_{21},\omega_{32},\omega_{30}$ with $\omega_{10}$ and $f-$factors. And one can see the prefactor 0.68 is due to the additional constraints. 

\subsection{Oscillator strength bound for other consecutive four-level combinations}
In this section, we consider the oscillator strength bound with transitions among arbitrary four consecutive levels $n-3$ through $n$, and the bound is shown to scale as $n^2$.

To start with, we first consider the transition among level 1, 2, 3 and 4, denoted as $h_4=x_{12}x_{23}x_{34}x_{41}$, and take the TRK sum rules with $(p,q)=(0,0),(1,1),(2,2),(3,3)$:
\begin{equation}
    \begin{aligned}
    &\omega_{10}\abs{x_{10}}^2 + \omega_{20}\abs{x_{20}}^2 + \omega_{30}\abs{x_{30}}^2 + \omega_{40}\abs{x_{40}}^2 + \sum_{n\geq5}\omega_{n0}\abs{x_{n0}}^2 = \frac{\hbar}{2m_e} \\
    &-\omega_{10}\abs{x_{10}}^2 + \omega_{21}\abs{x_{21}}^2 + \omega_{31}\abs{x_{31}}^2 + \omega_{41}\abs{x_{41}}^2 + \sum_{n\geq5}\omega_{n1}\abs{x_{n1}}^2= \frac{\hbar}{2m_e} \\
    &-\omega_{20}\abs{x_{20}}^2 - \omega_{21}\abs{x_{21}}^2 + \omega_{32}\abs{x_{32}}^2 + \omega_{42}\abs{x_{42}}^2 + \sum_{n\geq5}\omega_{n2}\abs{x_{n2}}^2= \frac{\hbar}{2m_e} \\
    &-\omega_{30}\abs{x_{30}}^2 - \omega_{31}\abs{x_{31}}^2 - \omega_{32}\abs{x_{32}}^2 + \omega_{43}\abs{x_{43}}^2 + \sum_{n\geq5}\omega_{n3}\abs{x_{n3}}^2= \frac{\hbar}{2m_e} 
\end{aligned}
\end{equation}
By substituting $x_{23}$, $x_{34}$ and $x_{41}$ with $x_{01}$ and $x_{12}$, we have:
\begin{equation*}
    \begin{aligned}
    &\abs{x_{23}} ={\frac{1}{\sqrt{\omega_{32}}}}\sqrt{\frac{\hbar}{m_e} - \omega_{10}\abs{x_{10}}^2 + \omega_{21}\abs{x_{21}}^2 - \sum_{n\geq3}\omega_{n0}\abs{x_{n0}}^2 - \sum_{n\geq4}\omega_{n2}\abs{x_{n2}}^2 }  \\
    &\abs{x_{34}} ={\frac{1}{\sqrt{\omega_{43}}}}\sqrt{\frac{3\hbar}{2m_e} - \omega_{10}\abs{x_{10}}^2 + \omega_{21}\abs{x_{21}}^2 + \omega_{31}\abs{x_{31}}^2 - \sum_{n\geq4}\omega_{n0}\abs{x_{n0}}^2 - \sum_{n\geq4}\omega_{n2}\abs{x_{n2}}^2 - \sum_{n\geq5}\omega_{n3}\abs{x_{n3}}^2}  \\
    &\abs{x_{41}} ={\frac{1}{\sqrt{\omega_{41}}}}\sqrt{\frac{\hbar}{2m_e} + \omega_{10}\abs{x_{10}}^2 - \omega_{21}\abs{x_{21}}^2 - \omega_{31}\abs{x_{31}}^2 - \sum_{n\geq4}\omega_{n1}\abs{x_{n1}}^2 }.
\end{aligned}
\end{equation*}
The maximum $\abs{h_4}$ is obtained when:
\begin{align*}
    &\sum_{n\geq3}\omega_{n0}\abs{x_{n0}}^2=0 \\
    &\sum_{n\geq3}\omega_{n1}\abs{x_{n1}}^2=0 \\
    &\sum_{n\geq4}\omega_{n2}\abs{x_{n2}}^2=0 \\
    &\sum_{n\geq5}\omega_{n3}\abs{x_{n3}}^2=0 \\
    &\omega_{10}\abs{x_{10}}^2=\frac{\hbar}{2m_e} \\
    &\omega_{21}\abs{x_{21}}^2\approx 1.33 \cdot \frac{\hbar}{2m_e}
\end{align*}
which result the bound:
\begin{equation} 
\abs{h_4}=\abs{x_{12}x_{23}x_{34}x_{41}} \lessapprox \frac{2.63}{4} \left( \frac{\hbar}{m_e} \right)^{2} \frac{1}{\sqrt{\omega_{21}\omega_{32}\omega_{43}\omega_{41}}}.
\end{equation}
Since this bound requires that the dipole moment $x_{01}$ reaches it maximum: $\omega_{10}\abs{x_{01}}^2=\hbar/(2m_e)$, it is 2.63X higher than that of $\abs{h_3}=\abs{x_{01}x_{12}x_{23}x_{30}}$. Note that here we only take the numerical solution instead of the exact one, since the analytical expression is hard to solve.

Furthermore, the bound can be generalized to the transitions among level $n-3,n-2,n-1$ and $n$, which is denoted as $h_n = x_{n-3,n-2}x_{n-2,n-1}x_{n-1,n}x_{n,n-3}$. Taking the TRK sum rules with $(p,q)=(0,0),(1,1),\ldots,(n-1,n-1)$, denoting $\omega_{n-2,n-3}\abs{x_{n-2,n-3}}^2=t$, and properly tuning the dipole moments other lower levels, we can have:
\begin{align*}
    \abs{h_n}  
    \leq &\sqrt{t \left( \frac{\hbar}{2m_e}+t \right) \left( \frac{\hbar}{m_e} + t \right)\left( \frac{(n-2)\hbar}{2m_e} - t \right) }  \\ 
    & \times \frac{1}{\sqrt{\omega_{n-2,n-3}\omega_{n-1,n-2}\omega_{n,n-1}\omega_{n,n-3}}}  
\end{align*}
The bound is obtained when:
\begin{align*}
    \frac{d}{dt} \left(  t \left( \frac{\hbar}{2m_e}+t \right) \left( \frac{\hbar}{m_e} + t \right)\left( \frac{(n-2)\hbar}{m_e} - t \right) \right)=0.
\end{align*}
Again, this is hard to solve analytically, here we only provide an asymptotic analysis: for large $n$, the optimal $t$ scales as $t\propto n\hbar/(2m_e)$, then the original objective function can be approximated as:
\begin{align*}
    \abs{h_n} \propto \sqrt{t^3\left(\frac{n\hbar}{2m_e}-t\right)} \propto t^2 \left( \frac{\hbar}{2m_e} \right)^2.
\end{align*}
The key result of this section is a scaling law for $|h_n|$: \mybox{$|h_n|\sim n^{2}$}. Furthermore, similar to the bound of arbitrary 3 levels, \eqref{arb 3 level}, the bound of arbitrary 4 levels will have the largest scaling law for $n$, where $n$ is the index of the first transition level. Therefore, one can design QWs to have highest first transition level as possible, to get at most $\sim n^2$ enhancement of oscillator strength for resonant third-order processes. For example, if the QW is designed to produce a resonant response among level 9,10,11,12, then it can yield at most 39X higher oscillator strength than that of the lowest four states.

\subsection{Comparison of bounds with various designs}

In this section, we list all the data points of designed QWs and corresponding bounds for SHG (Table \ref{table:shg-x01x12x20}), SFG (Table \ref{table:sfg-x01x12x20}) and THG (Table \ref{table:thg-x01x12x23x30}), as shown in the Fig. \ref{fig:shg bound}(c), \ref{fig:sfg bound}(b) and \ref{fig:thg bound}(b). For each data point, we summarized the designed QW properties, including reference, material, effective mass, energy spacing, oscillator strength, corresponding bound and the ratio between design and bound. In Table \ref{table:shg-x01x12x20} and \ref{table:thg-x01x12x23x30}, we only give the energy spacing between ground and first state $E_{10}$, since $E_{21}$ (and $E_{32}$) has a similar value with $E_{10}$. While in Table \ref{table:sfg-x01x12x20}, we list all the energy spacing values $E_{10}$, $E_{21}$. Data points with parentheses in the ``design/bound'' column are the theoretical design results, and corresponding design methods are labeled 
Theoretical designs are those data points with parentheses, labeling corresponding design methods in the ``design/bound'' column. These design methods with abbreviations are: ``Variational'': variational calculus method; ``SUSYQM'': supersymmetric quantum mechanics method; ``Discrete step'': discrete step QWs with finite barrier; ``IST'': inverse spectral theory; ``Superlattice'': Bragg-confined structures.


\begin{table}[h!] 
\centering
\begin{tabular}{||c c c c c c c||} 
 \hline
 References & QW material & $m_e^* (m_e)$ & $E_{10}$(meV) & $x_{01}x_{12}x_{20}$ (\SI{}{nm^3}) & bound (\SI{}{nm^3}) & \multicolumn{1}{p{2.5cm}||}{\centering design/bound}\\ [0.5ex] 
 \hline
 \cite{lee2014giant} & In$_{0.53}$Ga$_{0.47}$As & 0.0414 & 153 & 2.58 & 6.47 & 0.40 \\ 
 \cite{lee2016ultrathin} & In$_{0.53}$Ga$_{0.47}$As & 0.0414 & 150 & 3.57 & 9.06 & 0.39 \\
 \cite{kim2020spin} & In$_{0.53}$Ga$_{0.47}$As & 0.0414 & 127 & 2.72 & 8.27 & 0.33 \\
 \cite{sarma2019broadband} & In$_{0.53}$Ga$_{0.47}$As & 0.0414 & 116 & 4.08 & 9.44 & 0.43 \\ 
 \cite{rosencher1989second} & GaAs & 0.063 & 114 & 2.39 & 5.19 & 0.46 \\
 \cite{capasso1994coupled} & In$_{0.53}$Ga$_{0.47}$As & 0.0414 & 136 & 0.24 & 9.27 & 2.23 \\
 \cite{indjin1998optimization} & GaN & 0.18 & 240 & 0.30 & 0.36 & 0.83 (SUSYQM) \\
 \cite{indjin1998optimization} & GaN & 0.18 & 240 & 0.25 & 0.36 & 0.68 (Discrete step) \\
 \cite{radovanovic1999resonant} & GaAs & 0.067 & 240 & 0.48 & 1.60 & 0.30 (Superlattice) \\
 \cite{radovanovic2001two} & GaAs & 0.067 & 116 & 3.84 & 4.76 & 0.81 (SUSYQM) \\
 \cite{radovanovic2001two} & GaAs & 0.067 & 116 & 4.69 & 4.76 & 0.98 (Variational) \\
 \cite{tomic1998quantum} & GaAs & 0.067 & 100 & 4.38 & 5.95 & 0.74 (IST) \\
 \cite{tomic1998quantum} & GaAs & 0.067 & 116 & 3.91 & 4.76 & 0.82 (IST) \\
 \cite{radovanovic2004quantum} & GaN & 0.18 & 240 & 0.26 & 0.31 & 0.86 (SUSYQM) \\
 \cite{rosencher1991model} & GaAs & 0.067 & 100 & 1.20 & 1.35 & 0.89 (Discrete step) \\
 \hline
\end{tabular}
\caption{Comparison between designed oscillator strength and bounds of SHG, \eqref{bound-shg}.}
\label{table:shg-x01x12x20}
\end{table}

\begin{table}[h!] 
\centering
\begin{tabular}{||c c c c c c c||} 
 \hline
 References & QW material & $m_e^* (m_e)$ & $E_{10},E_{21}$(meV) & $x_{01}x_{12}x_{20}$ (\SI{}{nm^3}) & bound (\SI{}{nm^3}) & \multicolumn{1}{p{2.5cm}||}{\centering design/bound}\\ [0.5ex] 
 \hline
 \cite{fejer1989observation} & GaAs & 0.063 & 114,190 & 0.39 & 3.59 & 0.11 \\ 
 \cite{nefedkin2023overcoming} & In$_{0.53}$Ga$_{0.47}$As & 0.0414 & 453,162 & 0.38 & 2.58 & 0.15 \\
 \cite{nefedkin2023overcoming} & In$_{0.53}$Ga$_{0.47}$As & 0.0414 & 156,459 & 0.40 & 2.61 & 0.15 \\
 \cite{yildirim2017second} & CdSe & 0.119 & 242,203 & 0.13 & 0.77 & 0.17 (Discrete step) \\ 
 \cite{yildirim2017second} & CdSe & 0.119 & 280,223 & 0.12 & 0.63 & 0.20 (Discrete step) \\
 \hline
\end{tabular}
\caption{Comparison between designed oscillator strength and bounds of SFG, \eqref{bound-sfg}.}
\label{table:sfg-x01x12x20}
\end{table}

\begin{table}[h!] 
\centering
\begin{tabular}{||c c c c c c c||} 
 \hline
 References & QW material & $m_e^* (m_e)$ & $E_{10}$(meV) & $x_{01}x_{12}x_{23}x_{30}$ (\SI{}{nm^4}) & bound (\SI{}{nm^4}) & \multicolumn{1}{p{2.5cm}||}{\centering design/bound}\\ [0.5ex] 
 \hline
 \cite{yu2019third} & In$_{0.53}$Ga$_{0.47}$As & 0.0414 & 141 & 6.65 & 17.27 & 0.38 \\ 
 \cite{kim2020spin} & In$_{0.53}$Ga$_{0.47}$As & 0.0414 & 127 & 6.31 & 22.68 & 0.28 \\
 \cite{capasso1994coupled} & In$_{0.53}$Ga$_{0.47}$As & 0.0414 & 119 & 5.74 & 24.81 & 0.23 \\
 \cite{zahedi2019design} & In$_{0.53}$Ga$_{0.47}$As & 0.0414 & 124 & 9.32 & 21.72 & 0.43 \\ 
 \cite{radovanovic2001global} & GaAs & 0.067 & 116 & 8.24 & 9.47 & 0.87 (Variational) \\
 \cite{indjin1998optimization} & GaAs & 0.067 & 116 & 7.86 & 9.47 & 0.83 (SUSYQM) \\
 \cite{indjin1998optimization} & GaAs & 0.067 & 116 & 4.71 & 9.47 & 0.50 (Discrete step) \\
 \cite{indjin1998optimization} & GaAs & 0.067 & 116 & 2.43 & 9.47 & 0.26 (Discrete step) \\
 \cite{radovanovic1999resonant} & GaAs & 0.067 & 116 & 4.42 & 9.47 & 0.47 (Superlattice) \\
 \hline
\end{tabular}
\caption{Comparison between designed oscillator strength of QWs and bounds of THG, \eqref{bound-thg}.}
\label{table:thg-x01x12x23x30}
\end{table}

\clearpage

\bibliography{main}


\end{document}